\documentclass[aps,prl,reprint,longbibliography,superscriptaddress]{revtex4-1}

\pdfoutput=1

\usepackage{graphicx}  % needed for figures
\usepackage{dblfloatfix}
\usepackage{dcolumn}% Align table columns on decimal point
\usepackage{bm}% bold math
\usepackage{amsmath}
\usepackage{graphicx, , setspace, epstopdf, fancyhdr, mathrsfs, multirow}
\usepackage{amssymb}
\usepackage{color}
\usepackage{soul}
\usepackage[normalem]{ulem}

\usepackage[breaklinks,colorlinks=true]{hyperref}

\newcommand{\spvec}[1]{\ensuremath{\mathbf{#1}}}

\newcommand{\unitvec}[1]{\hat{\mathbf{{#1}}}}

\newcommand{\EQREF}[1]{Eq.~(\ref{#1})}
\newcommand{\bea}{\begin{eqnarray}}
\newcommand{\eea}{\end{eqnarray}}
\newcommand{\beq}{\begin{equation}}
\newcommand{\eeq}{\end{equation}}

\newcommand{\stochx}{\ensuremath{X}}
\newcommand{\stochxv}{\spvec{\stochx}}

\newcommand{\rv}{{\bf r}}

\newcommand{\Dc}{{\cal D}}

\renewcommand{\>}{\rangle}

\newcommand{\commentout}[1]{{}}

\newcommand{\av}[1]{\ensuremath{\langle #1 \rangle}}

\begin{document}

\title{Collective suppression of optical hyperfine pumping in dense clouds of atoms in microtraps}

\author{Shimon\ Machluf}
\affiliation{Van der Waals-Zeeman Institute, Institute of Physics, University of Amsterdam, Science Park 904, 1098XH Amsterdam, The Netherlands}
\author{Julian B.\ Naber}
\affiliation{Van der Waals-Zeeman Institute, Institute of Physics, University of Amsterdam, Science Park 904, 1098XH Amsterdam, The Netherlands}
\author{Maarten L.\ Soudijn}
\affiliation{Van der Waals-Zeeman Institute, Institute of Physics, University of Amsterdam, Science Park 904, 1098XH Amsterdam, The Netherlands}
\author{Janne Ruostekoski}
\affiliation{Department of Physics, Lancaster University, Lancaster, LA1 4YB, United Kingdom}
\author{Robert J.\ C.\ Spreeuw}
\affiliation{Van der Waals-Zeeman Institute, Institute of Physics, University of Amsterdam, Science Park 904, 1098XH Amsterdam, The Netherlands}

%\date{\today}

\begin{abstract}
We observe a density-dependent collective suppression of optical pumping between the hyperfine ground states in an array of submicrometer-sized clouds of dense and cold rubidium atoms. The suppressed Raman transition rate can be explained by
strong resonant dipole-dipole interactions that are enhanced by increasing atom density, and are already significant  at densities of $\alt 0.1 k^3$, where $k$ denotes the resonance wavenumber. The observations are consistent with stochastic electrodynamics simulations that incorporate the effects of population transfer via internal
atomic levels embedded in a coupled-dipole model.
\end{abstract}

\maketitle

Interfaces between trapped atoms and light play a central role, e.g., in sensing~\cite{BudkerRamalisNatPhys2007}, metrology~\cite{RevModPhys.87.637},
nonlinear devices~\cite{HarrisPhysToday1997,FleischhauerEtAlRMP2005}, and quantum information processing~\cite{HAM10}. For example, trapped cold atomic ensembles are utilized in the development of quantum memories, quantum repeaters, and as an interface to convert between flying and standing qubits \cite{Duan2001Longdistance,Matsukevich663,vanderWal2003Atomic,ChanelierePhotStorNAT2005,Chou2005Measurementinduced,Chou2007Functional,Yuan2008Experimental}.  For many quantum information protocols, such as those based on Rydberg interaction \cite{Saffman:RevModPhy2010, Dudin2012Observation}, it is also essential to confine the atoms in lattices~\cite{SchaussSci2015, EndresSci2016, BarredoSci2016, ZeiherNatPhy2016}.
A typical length scale of Rydberg dipole-dipole (DD) interaction is around $5\,\mu$m, necessitating a comparable size lattice spacing and even smaller individual trap size.
Engineering  smaller lattice periods attract interest as a route towards interaction between atoms and nanostructured surfaces \cite{Thompson2013Coupling} or increased tunneling rates of atoms between adjacent sites~\cite{Bloch2005Ultracold, Leung2011Microtrap, Gullans2012Nanoplasmonic, Wang2017Trapping, Romero-Isart2013Superconducting}.
For example, in order to increase the hopping energy
above the cloud temperature one would benefit from lattice spacings well below the optical wavelength~\cite{YiNJP2008}.

Understanding the fundamental properties of the interaction of resonant light with trapped  atomic ensembles is essential for all of the above applications, and has consequently attracted considerable recent experimental interest~\cite{BalikEtAl2013,Pellegrino2014a,Cha14,wilkowski2,Ye2016,Guerin_subr16,vdStraten16,Keaveney2012,Jenkins_thermshift,Jennewein_trans,Dalibard_slab}. The observed phenomena in atomic systems include suppression of light scattering in small samples~\cite{Pellegrino2014a}, subradiance~\cite{Guerin_subr16},  superflash effects~\cite{wilkowski2}, and  the dependence of the response on the quantum statistics~\cite{vdStraten16}. Several established models of the resonance response in continuous medium electrodynamics may be violated in cold and dense atomic ensembles~\cite{Jenkins_thermshift,Jennewein_trans,Javanainen2014a,JavanainenMFT}.
This is because each atom is subject to the driving field plus the fields scattered by the other atoms.
These fields mediate strong resonant DD interactions between the atoms, resulting in collective excitations whose behavior cannot be  described as a sum of the responses of isolated atoms.

Most experiments on the collective optical responses of cold atoms have been performed either in dilute ensembles \cite{sheremet2014} or have focused on elastically scattered light (in terms of the internal atomic state), where the contribution from optical pumping has not
been measured~\cite{Pellegrino2014a}. These can frequently also be theoretically analyzed by coupled-dipole model simulations~\cite{Javanainen1999a} in the limit of low light intensity on the assumption that  each atom responds to  light as a simple damped linear harmonic oscillator.
Here we extend both the experiment and theory, and report on strong \emph{collective}, \emph{density-dependent} suppression of optical  pumping between hyperfine levels, due to resonant DD interactions in magnetically trapped submicron clouds.
This is relevant for many applications of atom-light interfaces, e.g., for the optical protocols of quantum memories for storing photons via spontaneous Raman scattering~\cite{Duan2001Longdistance}.
We perform our measurements in a lattice of magnetic microtraps on an atom chip~\cite{Keil:2016wt,Leung2014Magneticfilm}.
Microtraps provide a natural platform with cold-atom-light interfaces at high atom densities and large optical cross-sections~\cite{WhitlockNJP2009}
in a structured environment, where parallel measurements over a large range of initial densities are simultaneously obtained at each experimental run.

In the experiment, optical pumping is performed between hyperfine ground states of  dense, submicrometer-sized rubidium ensembles and measured by detecting the remaining atoms in the initial state.
Standard linear coupled-dipole model simulations neglect the atomic levels and cannot model optical pumping. In the theoretical analysis, we therefore implement stochastic electrodynamics simulations  based on a recently proposed model of coupled many-atom internal level dynamics~\cite{Lee16} that incorporate the population transfer between atomic levels.
Simulations and experiment qualitatively agree and reveal a collective density-dependent suppression
of optical pumping; the Raman transitions between the internal levels are suppressed due to strong many-body resonant DD interactions.
We also numerically identify a collective transition resonance that is blue-shifted as the density is increased.

\begin{figure}[t!]
\centering
\includegraphics[width=\columnwidth]{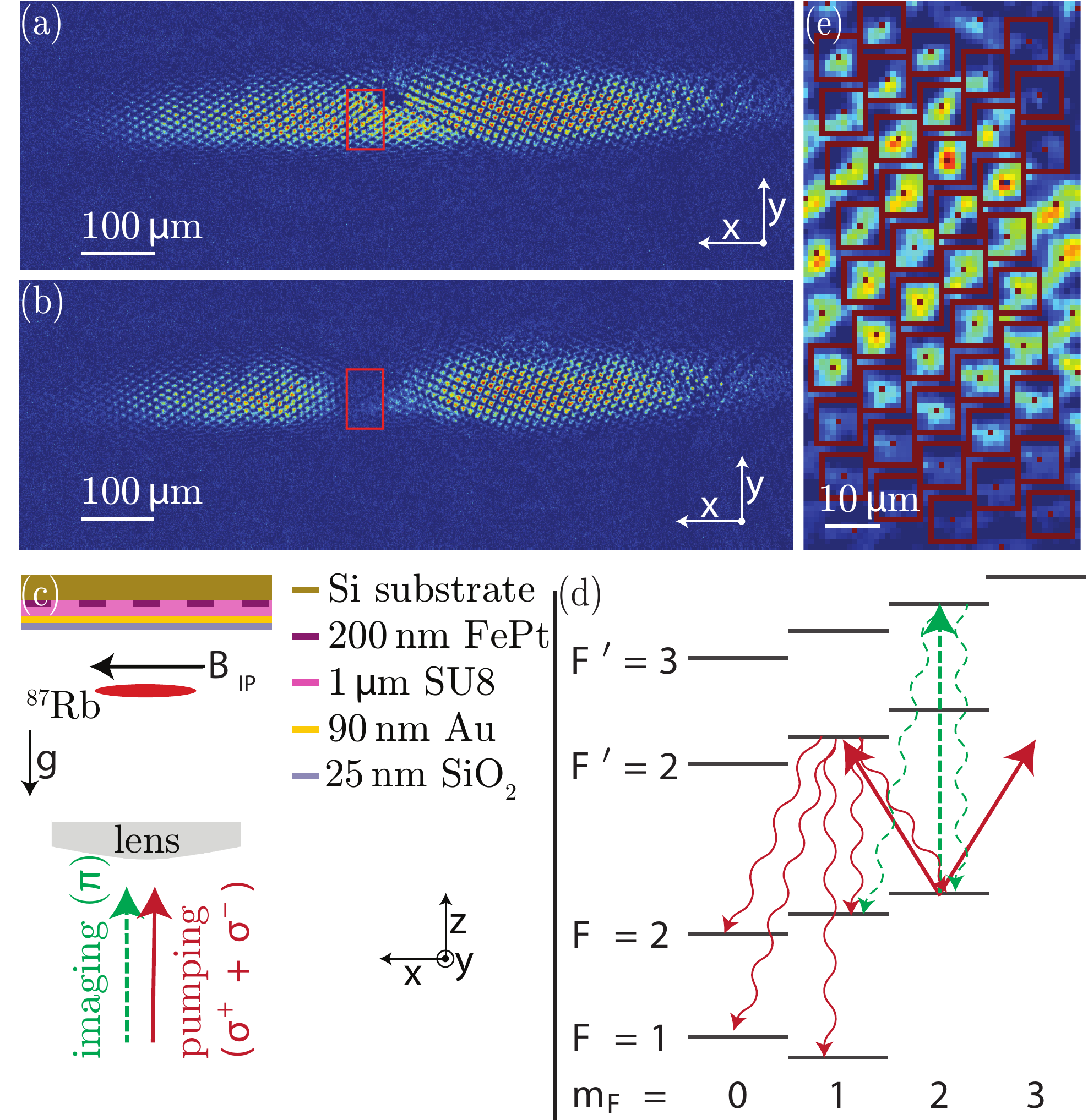}
\caption[]{Description of the  apparatus.
(a) A sample image after the macroscopic cloud is loaded into the lattice of the microtraps.
The false color indicates the integrated atomic density along the imaging axis (optical density).
(b) Similar to (a), but with an area of visibly depleted microtraps after a long laser pulse ($>100\,\mu$s).
(c) A sketch of the apparatus showing the atom chip (with its different layers), the macroscopic cloud below the atom chip, the lens used for imaging, and the two relevant lasers: pumping and imaging.
(d) A sketch of the relevant atomic levels (hyperfine structure and Zeeman sublevel).
The straight arrows show the first possible excitation from the $|2,2\rangle$ state, and the wavy arrows show the decay channels.
The reabsorbed photons have many more absorption and emission paths, which are not shown.
(e) A magnification of the area marked by the red rectangle in (a) showing the individual microtraps.
Each trap's center is marked with a dot, and the squares mark the area where atoms are counted and considered to belong to one microtrap.
}
\label{fig:ODImages}
\end{figure}

The apparatus is described in~\cite{Leung2014Magneticfilm}.
Briefly, we have an atom chip  where the trapping potential is generated by a patterned layer of permanent magnet  (FePt) film.
Together with a homogeneous magnetic field, a triangular lattice (with $10\,\mu$m spacing) of Ioffe-Pritchard type microtraps is created $\sim 8\,\mu$m from the surface.
We load $^{87}$Rb atoms in the  (5$S_{1/2}$) $|F,m_F\rangle =|2,2\rangle$ state at  $\sim 15\,\mu$K temperature into the microtraps [see Fig.~\ref{fig:ODImages}(a) for a sample image] with calculated averaged trapping frequency of $\bar{\omega}=2\pi\times14\,$kHz [(7,18,22)$\,$kHz in the $(x,y,z)$ directions]. Separate measurements show that the temperature is approximately uniform across the lattice, with variations at the level of the measurement accuracy, $\sim 2\mu$K. Therefore, the
calculated trap size (root-mean-square of the density, $\sigma_i=\sqrt{k_BT/m\omega_i^2}$) is independent of the number of atoms, $\sigma_{x,y,z}=(0.86, 0.34, 0.27)\,\mu\text{m}=(6.9,  2.7, 2.2)\,1/k$. The peak atomic density is $\rho=8\times10^{11}\,N/$cm$^{3}=0.0015\,N \times k^3$, with $N$ the number of atoms in the microtrap and $k=2\pi/\lambda$ ($\lambda=780$nm the transition wavelength).

After loading we wait $\sim20\,$ms for untrapped atoms to leave the microtraps area.
We then pulse a focused laser beam ($\sim100\,\mu$m  waist) with detuning $\Delta=\Omega-\omega$ (the laser and atomic transition frequencies, respectively) from the (5$P_{3/2}$) $F'=2$  transition \cite{Naber:2015wk} for a varying length of pumping time $t_\text{p}$, causing atom loss due to decay to untrappable or dark states [Fig.~\ref{fig:ODImages}(b)]; see Fig.~\ref{fig:ODImages}(c,d) for a sketch of the apparatus and the induced transitions.
This pumping laser has  $\sigma^+ + \sigma^-$ polarization and a power of $\approx13\,\text{nW}$, which results in a saturation parameter of $s\approx 4 \times10^{-3}$ for the pumping laser transition. We neglect the spatial variation of $s$ ($\pm10$\%) over the analyzed sites.

After the pulse we detect the remaining atoms with absorption imaging via the $F=2\rightarrow F'=3$ transition with  $\pi$ polarized light. We use optical Bloch equations (OBE) to calculate the atom numbers from the effective absorption cross section during the probe pulse.
We also remove noise from the image using a fringe removal algorithm and deconvolute the image with an experimentally measured point-spread-function~\cite{SOM}.
The analysis of the images~\cite{SOM} is done on a region marked with the red rectangle in Fig.~\ref{fig:ODImages}(a,b), and magnified in Fig.~\ref{fig:ODImages}(e).

To reduce the experimental noise, we sort the microtraps  into 9 groups based on their initial number of atoms~\cite{SOM}.
%These groups are the basis for the analysis.
The difference in initial population is due to the Gaussian shape of the original atom cloud.

\begin{figure}[t]
%\centering
\includegraphics[width=\columnwidth]{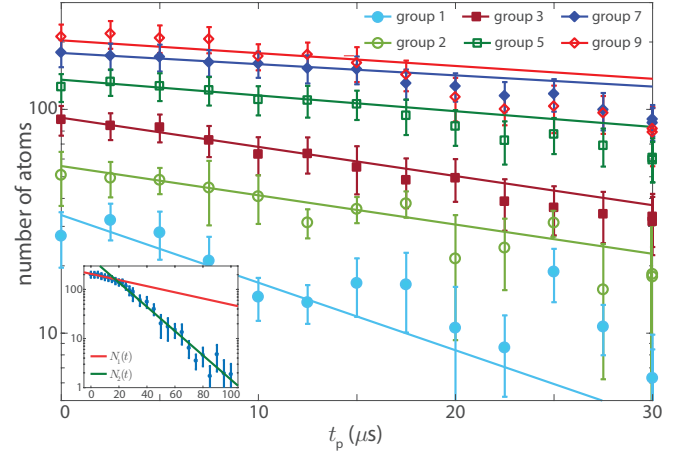}
\caption{Number of atoms as a function of pumping  times (at $\Delta=0$) in a semilog plot for different groups of microtraps, representing different initial densities.
Initial density increases from group 1 (lower curve) to group 9 (upper curve), and for clarity not all nine groups are shown. For
$t<15\,\mu$s, each data set is fitted to an exponent that decreases as the density increases.
For $t>15\,\mu$s, we use a different exponent $\gamma_{2}$ that shows faster decay (for group 8) in the inset.
The errorbars (root-mean-square) are due to repetitions and distribution of atom numbers within each group of microtraps, and the calculated absorption cross section.
}
\label{fig:ExpData1}
\end{figure}

Figure \ref{fig:ExpData1} shows the decay of initial state population, as the atoms are pumped to dark or untrapped states, with each curve representing a different group of microtraps.
We see a change of the slope after $t_\text{p}\approx 15\,\mu$s (clearly visible in the inset), and, hence, fit the data to two separate exponents, $N_{1,2}^0\,\text{exp}(-\gamma_{1,2} t)$, where the $1,2$ indices refer to $t_\text{p}\leq15\,\mu$s and $t_\text{p}>15\,\mu$s, respectively.
Importantly, the transition between the two exponents occurs at a fixed time, rather than at a fixed density. This indicates that the second exponent is an effect of our measurement procedure
and may be partly caused by multilevel effects (untrapped sublevels leave the trap region in $10-15\,\mu$s), as well as light-induced dipolar forces and collisions that were also shown to lead to enhanced losses at
longer times in Ref.~\cite{Pellegrino2014a}.

\begin{figure}[t]
\includegraphics[width=\columnwidth]{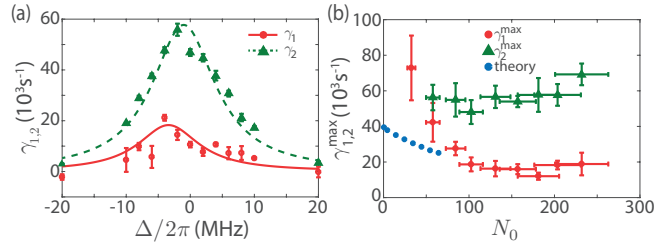}
\caption{(a) The pumping rates obtained from the two exponents $\gamma_{1,2}$ for different detunings and $s=4\times10^{-3}$.
The fitted Lorentzian (from group 8) for $t<15\,\mu$s is clearly suppressed (in amplitude) compared with the case $t>15\,\mu$s.
(b) Density-dependent suppression of the pumping rate.
The maximum  rate [$\gamma^\text{max}_{1,2}$ in (a)] is plotted as a function of the averaged initial number of atoms $N_0$ of each group (giving the peak density $\rho/k^3=0.0015N_0$). In the least populated traps, group 1, the data point of $\gamma^\text{max}_2$
is absent due to low number of atoms that is below our noise level. We also show the results of numerical simulations of
stochastic electrodynamics for the same values of $s$ and the trap size as in the experiment.
}
\label{fig:ExpData2}
\end{figure}

Figure \ref{fig:ExpData2}(a) shows the fitted decay rates, $\gamma_{1,2}$, as a function of the pumping laser detuning, forming a Lorentzian shape transition rate, with fitted maximum decay rates $\gamma^\text{max}_{1,2}$.
Figure \ref{fig:ExpData2}(b) summarizes the results for the different groups (i.e., different initial densities) of microtraps.
The amplitudes of the  fitted Lorentzians  are shown for   $t_\text{p}\le15\,\mu$s ($\gamma_1^\text{max}$) and for $t_\text{p}>15\,\mu$s ($\gamma_2^\text{max}$).
The density dependence is only visible at short pumping times, i.e., in $\gamma_1^\text{max}$. At high densities, $N_0\agt 100$, the suppression is strong, by up to a factor four.
Within our signal-to-noise level, we do not observe any density-dependence of the Lorentzian width or shift.

The density-dependent suppression of the pumping rate $\gamma_1^\text{max}$ constitutes the central result of the paper. The suppression cannot be explained by the independent-atom OBE but results from the collective response of the atoms, generated by the strong resonant DD interactions.
We also checked and ruled out several alternative explanations for the suppression.
For example, the size of the traps does not depend on the number of atoms. This is because the atoms obey the Boltzmann distribution, and the cloud size is determined by the trap frequency and temperature. The trap frequency depends on the precision of the lithographic patterning of the magnetic film, with negligible errors at the length scale of 10$\mu$m. The temperature is defined by the forced evaporative cooling stage that we use after loading the traps. The final temperature is determined by the trap depth at the end of the evaporation ramp, which is the same for all the traps, yielding approximately uniform temperature across the lattice. We also rule out the effects of inhomogeneous broadening due to finite temperature and Zeeman shifts as they are a few 100kHz, much less than the natural linewidth or the observed broadening.

We find that significant suppression starts at surprisingly low atom densities of $\rho/k^3\alt 0.1$, which is especially relevant
for the operation of quantum
devices~\cite{BudkerRamalisNatPhys2007,RevModPhys.87.637,HarrisPhysToday1997,Matsukevich663,vanderWal2003Atomic,Yuan2008Experimental,Saffman:RevModPhy2010}
and
protocols~\cite{Dudin2012Observation, Leung2011Microtrap, Gullans2012Nanoplasmonic, Wang2017Trapping, Romero-Isart2013Superconducting} that rely on the interaction between light and trapped atoms.
For example, communication protocols~\cite{Duan2001Longdistance} are based on spontaneous Raman scattering, and as the atomic systems become smaller and denser, this rate will be suppressed.

\begin{figure}[t]
\centering
\includegraphics[width=\columnwidth]{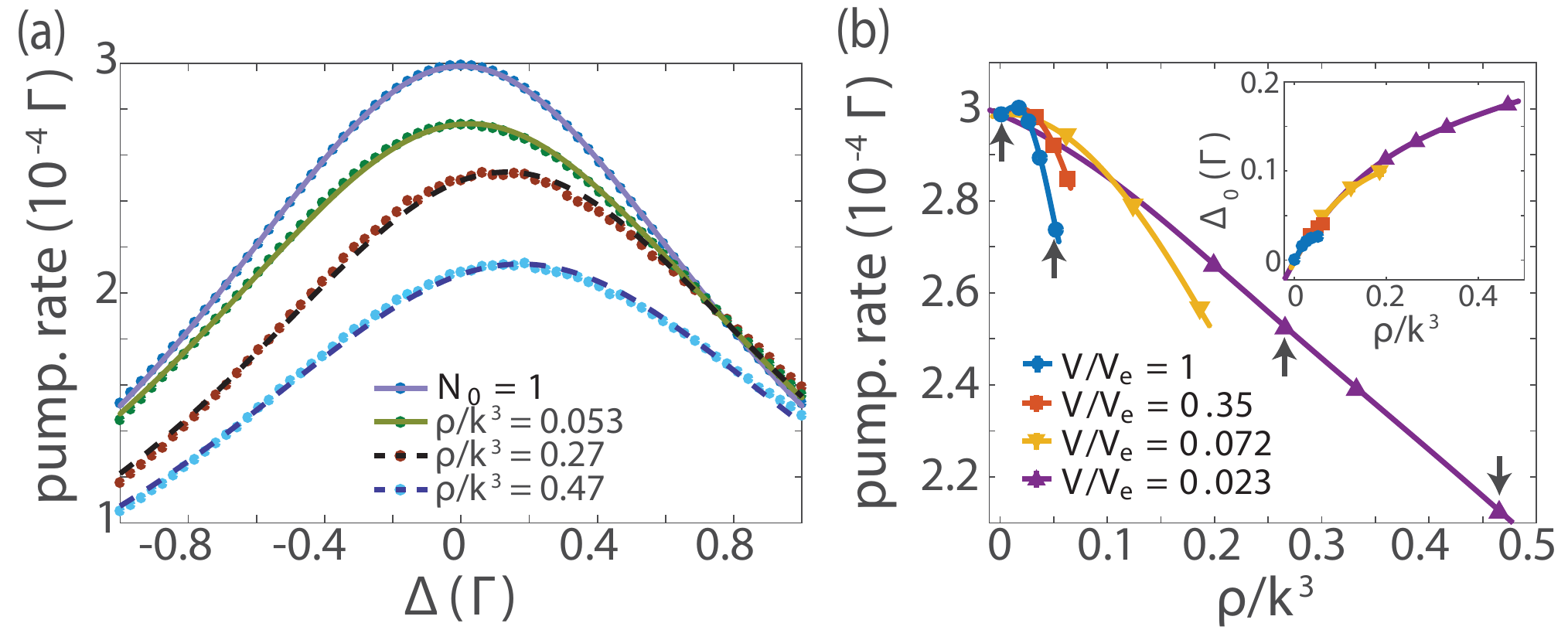}
\caption{Stochastic simulations of density-dependent optical pumping suppression due to collective DD interactions in a three-level system for $s=6\times 10^{-4}$ and the experimental trap aspect ratio $\sigma_{x,y,z}=(6.9,  2.7, 2.2)\,1/k$.
(a) The pumping rate per atom as a function of the detuning of light from the single-atom resonance with increasing initial density $\rho$ for the curves from top to bottom.
The simulation results (dots)  are shown with lines denoting fitted Lorentzians for the experimental volume $V= V_e$ (solid line) and $V=0.023V_e$ (dashed line).
(b) The fitted amplitudes and resonance shifts (inset) of Lorentzians as in (a).
The simulations for the different traps are performed for up to $N_0=35, 15, 9, 7$ atoms (for $V/V_e=1, 0.35, 0.072, 0.023$, respectively)~\cite{SOM}.
The solid lines are interpolations.
The points marked with arrows are the fitted amplitudes of the four data-sets shown in (a).
}
\label{fig:theory}
\end{figure}

Along with the experimental observations we performed stochastic electrodynamics simulations that show the collective density-dependent suppression of the optical pumping in a qualitative agreement with the
experiment, see Figs.\ \ref{fig:ExpData2} and \ref{fig:theory}.
In the coupled-dipole model simulations~\cite{Javanainen1999a} the electrodynamics of radiatively coupled
atoms is solved for stochastically sampled atomic positions from the density distribution.  Ensemble-averaging over many
such realizations then yields the optical response.
The stochastic treatment of atomic coordinates establishes
the position-dependent correlations between the atoms that go beyond the standard mean-field theory of continuous medium electrodynamics~\cite{JavanainenMFT}.
In the limit of low light intensity the coupled-dipole model simulations modeling stationary (cold) atoms with only one electronic ground state are exact~\cite{Lee16}, and for laser-cooled Rb atoms
thermal motion on the timescale of the multiple scattering process of a single photon is negligible, while the spatial averaging compensates the atomic motion during the pulse \cite{motion}.
Here we extend the standard coupled-dipole model, that neglects the atomic levels and treats the atoms as oscillating dipoles,  using a recently proposed model~\cite{Lee16} of coupled many-body internal atomic level dynamics that incorporates the
effects of population transfer. We do this by introducing a semiclassical approximation that neglects the quantum entanglement between the internal
electronic levels and allows significantly larger atom numbers than full quantum treatments~\cite{Olmos16}. Closely related semiclassical approaches have also been introduced in Refs.~\cite{Sutherland_satur,Lee17}.

The formalism is explained in detail in~\cite{SOM}. In each stochastic realization of $N$ atomic positions ${\{\stochxv_1,\ldots,\stochxv_N\}}$, we write a single particle density matrix $\hat \rho_{a b}(\rv)$ for  the different electronic sublevels $a,b$ as
the sum over the atoms $j$,
$
\av{\hat \rho_{a b}(\rv)}_{\{\stochxv_1,\ldots,\stochxv_N\}} =
\sum_j \rho_{a b}^{(j)}\delta(\rv-\stochxv_j)
$.
Instead of considering the full experimental configuration of all the $F=1,2$ and $F'=2$ electronic levels, we approximate the system by an effective three-level model where
one of the ground levels refers to the initial state $|1\>\equiv|2,2\>$, and all the final electronic ground levels are approximated by a single state $|2\>$.
Resonant incident light then drives the transition $|1\>\leftrightarrow |e\>$ to an electronically excited state $|e\>$, and the atoms can spontaneously decay to both levels $|1\>$ and $|2\>$.
For the equal transition strengths for the two levels, a set of coupled equations of motion for internal level one-body density matrix elements $\rho_{a b}^{(j)}$ ($a,b=1,2,e$), for each atom $j=1,\ldots,N$ then take a simple form.
For example,
\begin{align}
{d\over dt}\,&\rho_{1 1}^{(j)} =
+\Gamma  \rho_{e e}^{(j)} +\sqrt{2}{\rm Im} \big[\frac{\xi}{\Dc}\rho_{e 1}^{(j)}
  \unitvec{e}^\ast_1 \cdot
{\bf D}^{+}_F(\stochxv_j)\big] \nonumber \\
&+ {\rm Im} \big[ {\xi} \sum_{l\neq j}\,
 \mathcal{G}^{(jl)}_{1g}  \rho_{g e}^{(l)}
\rho_{e 1}^{(j)} \big] \,.
\label{eq:drhog_fullstochastic}
\end{align}
   Here the summations run over the $N$ atoms and the ground levels $g=1,2$; $\Dc $ is the reduced dipole matrix element, $\xi=\Dc^2/(\hbar\epsilon_0)$, $ \unitvec{e}_1$ is the unit vector along the direction of the dipole matrix element for the $|1\>\leftrightarrow |e\>$ transition, and  $\Gamma$ denotes the half-width at half maximum (HWHM) resonance linewidth.
We treat the positive frequency component of the slowly-varying incident light field ${\bf D}^{+}_F$ as a plane wave. The last term in \EQREF{eq:drhog_fullstochastic} describes the light-mediated interactions between
the atoms $j$ and $l$, where $\mathcal{G}^{(jl)}_{g g'}$ denotes the dipole radiation from the $g'\leftrightarrow e$ transition of the atom $l$ to the $g\leftrightarrow e$ transition of the atom $j$~\cite{SOM}.
In the absence of the coupling terms $\mathcal{G}^{(jl)}_{g g'} $, the equations reduce to  OBE.
The terms $\mathcal{G}^{(jl)}_{g g'} $ represent the strong resonant DD interactions that depend on the relative positions between the atoms and lead to spatial
correlations in the optical response.

The suppression of the pumping rate in the simulations~\cite{SOM} is illustrated in Fig.~\ref{fig:theory} for different atom numbers  and trap sizes.
The $N_0=1$
result represents the solution that is obtained by solving OBE. Consequently, the suppressed pumping rates per atom
for the higher densities are a direct consequence of the collective resonance DD interactions between the atoms.
Although simulations using the full range of experimental atom numbers are not feasible, for $N_0\leq65$ we find qualitatively similar behavior due to the collective density-dependent effects, as shown in Fig.~\ref{fig:ExpData2}(b).
The decay is slower in the simulations than in the experiment, which we attribute to the simplified level scheme. To illustrate this, when we  incorporate the full multilevel structure in the OBE calculation, we find a 60\% higher rate ($61\times 10^3$~s$^{-1}$) than the three-level OBE [$38\times 10^3$~s$^{-1}$; $N_0=1$ theory point in Fig.~\ref{fig:ExpData2}(b)].

In our experiment the variation of the cloud size between different measurements is negligible.
Even though we can therefore rule out that the observed suppression is due to size differences of the atom clouds, we numerically studied how the suppression is affected by the sample size [Fig.~\ref{fig:theory}(b)]. We found that for the same density the smaller traps are less suppressed. This indicates how pumping can be suppressed by both the increase in density and the increase in optical depth (increase in size) while keeping the density fixed.

The suppression can be understood by microscopic mechanisms.
As the level shifts generated by the DD interactions are sensitive to the relative atomic positions, each random configuration of the positions produces different shifts, effectively tuning the atoms off resonance by
different amounts and generating something reminiscent of inhomogeneous broadening. Moreover, the pumping can also be suppressed when the atoms decaying to final states transfer back by reabsorption of
photons.

The measurements of the resonance shifts in the spectroscopy of dense atom samples have attracted considerable attention recently~\cite{Keaveney2012,Jenkins_thermshift,Ye2016,Jennewein_trans,Dalibard_slab}, and especially the origin of the shifts (or the absence of them) has been actively studied~\cite{Javanainen2014a,JavanainenMFT,Jenkins_thermshift,Jennewein_trans,Dalibard_slab}.
Although we were not able to resolve them experimentally, the simulations in Fig.~\ref{fig:theory}(a) [and summarised in the inset in Fig.~\ref{fig:theory}(b)] show a blue-shifted collective resonance as the density is increased.
For the calculated cases, the density-dependence of the shift is no longer linear.
Moreover, it is about an order of magnitude less than the Lorentz-Lorenz shift and has the opposite sign, consistently with the recent transmission measurements~\cite{Dalibard_slab}.
Interestingly, we also find that the shift is larger for smaller traps at the same density, indicating dependence on the system size.

To conclude, we show experimentally and theoretically that optical pumping is suppressed in small, dense clouds due to collective resonant DD interactions.
The observed suppression, by up to a factor four, is already significant  at densities of $\rho/k^3\alt 0.1$.
In addition, the simulations show a collective transition resonance that is blue-shifted as the atom density is increased.

\acknowledgments
We thank C.\ Adams, T.\ Pfau, and M.\ D.\ Lee for stimulating discussions. Our work is financially supported by the Netherlands Organization for Scientific Research (NWO)
and EPSRC.
JBN acknowledges financial support by the Marie Curie program ITN-Coherence (265031).


\begin{thebibliography}{53}
\bibitem [{mot()}]{motion}%
  \BibitemOpen
  \href@noop {} {}\bibinfo {note} {For long pulses, different photons in the
  same pulse may probe different atomic positions due to the atomic motion not
  being negligible during the length of the pulse. We have checked the effect
  of this by randomizing the atomic positions also at fixed times during the
  pulse, and have found the difference in the population decay to be much less
  than 1\% \cite{SOM}.}\BibitemShut {Stop}%
\end{thebibliography}
\end{document}